\documentclass[twocolumn,english,aps,superscriptaddress,prb]{revtex4-2}

\usepackage{array}
\usepackage[latin9]{inputenc}
\usepackage{amsmath}
\usepackage{amssymb}
\usepackage{graphicx}
\usepackage{babel}
\usepackage{mathrsfs}
\usepackage{amsfonts}
\usepackage{epstopdf}
\usepackage{multirow}
\usepackage{color}
\usepackage{natbib}
\usepackage{bm}

\begin{document}

\title{Discrete global symmetries and dynamics of emergent fermions}
\author{Fan Yang}
\affiliation{Institute for Advanced Study, Tsinghua University, Beijing, China 100084}
\author{Fei Zhou}
\affiliation{Department of Physics and Astronomy, University of British Columbia, Vancouver, BC, Canada, V6T 1Z1}
\date{\today}

\begin{abstract}

Global symmetries that define the number of low energy degrees of freedom
have profound consequences on universal properties near topological quantum critical points and in other gapless or nearly gapless states of emergent fermions.
We take a $Z_2$ global symmetry (such as time-reversal) as an example to study its effect on thermodynamic and transport properties. Although the thermal entropy density of $Z_2$ symmetric systems is simply twice of their counterparts without any global symmetries or the $Z_1$ class, the temperature dependence of thermal conductivity $\kappa$ is distinctly and drastically different for different symmetries. 
For systems with dynamic exponent $z=1$, in the $Z_2$ symmetric class, we have $\kappa\propto T^{-(d-1)}$ in the quantum critical regime near weakly interacting fixed points, while for systems with no global symmetries (i.e., the $Z_1$ class), we have $\kappa\propto T^{-(d+3)}$, with $d$ being the spatial dimension. Only near strong coupling fixed points, both cases with or without $Z_2$ global symmetries follow the same scaling function, $\kappa \propto T^{d-1}$.
These distinct scalings of thermal conductivity can also appear in gapless surface Majorana states.
\end{abstract}

\maketitle

\section{Introduction}
Tremendous efforts have been made to understand emergent Majorana fermions in electronic systems \cite{Alicea12, Beenakker13,Elliott15}.  
Many previous studies on Majorana fermions in condensed matter have mainly focused on Majorana zero modes localized in vortices or on the edge of topological superconductors \cite{Read00,Kitaev01,Fu08,Qi10a,Lutchyn10,Sau10,Alicea10,Chung11,Cook11,Gangadharaiah11,Choy11}. One major motivation of those efforts is to trap and manipulate these decoherence free states as a fundamental building block for topological quantum computers \cite{Kitaev03,Nayak08}.
On the other hand, propagating Majorana fermions as low energy emergent particles have received relatively less attention, partly because they are hard to separate experimentally. 
The main purpose of this  article is to reveal some surprising collective dynamics of Majorana fermions that are distinct from complex fermions and can be potentially studied in experiments.

There are at least two different classes of phenomena where the dynamics can be characterized by the theoretical results presented in this  article. 
The first class of phenomena are boundary or surface dynamics in a topological state.
Around a typical topological bulk, surface fermions are gapless and protected by symmetries \cite{Hasan10,Qi11,Bernevig}. Many aspects of interaction dynamics on gapless surfaces can be related to quantum critical phenomena \cite{Sachdev}, as was discussed in Ref. \cite{Yang21}.
Consequently, its transport properties should resemble those in a quantum critical regime after a proper symmetry identification.

The other class is related to quantum critical regimes where universality classes are defined by emergent gapless {\em real} fermions, which also turn out to be topological in nature.
Generally, between quantum phases with the same local ordering but different topologies, there are always topological quantum critical points (TQCPs). These TQCPs are not induced by usual spontaneous symmetry breaking, and are beyond the standard Landau paradigm of order-disorder phase transitions. Instead, they appear to signify a change of global topologies. Depending on the topological phases involved \cite{Schnyder08,Kitaev09,Qi10,Teo10,Kennedy16, Chiu16}, TQCPs possess different discrete global symmetries or no symmetries at all.
A particularly interesting subset is the TQCPs in fermionic superfluids and superconductors \cite{Yang19,Yang21,Zhou22}. One unique aspect of superconductors is that particle and hole excitations are indistinguishable due to the presence of the condensate of Cooper pairs. 
Thus, TQCPs in superconductors are naturally characterized by dynamics of real fermions in the bulk \cite{Read00,Yang19,Yang21,Zhou22}.
Below we will mainly carry out our discussions in the context of TQCPs although one shall keep in mind that their main practical applications are to gapless Majorana fermions as surface or edge states that are experimentally more accessible. 

The key concept here is that it is {\em discrete global symmetries} at TQCPs that define the degrees of freedom at low energies \cite{Zhou22}. For example, if apart from the emergent particle-hole symmetry no other symmetries are present, generically the low energy effective field theory describing the TQCPs should only consist of a minimum $n=2$ or two-component real fermion fields in its fundamental representation. This case of no global symmetries can also be named as the class of $Z_1$ \cite{symmetries}. 

If there do exist discrete global symmetries, the effective field theory should consist of $n\geq4$-component real fermions. In particular, if only a $Z_2$ global symmetry such as time-reversal symmetry (TRS) is present, we have an $n=4$ real-fermion field representation for TQCPs. The differences in $n$, the degrees of freedom near TQCPs, dictate the structure of effective interactions, which in turn determines the universal behavior of thermodynamic and transport quantities in the quantum critical regime. 
In our studies, we have found two distinct classes of scaling behaviors of thermal transport: (1) the $Z_1$ class without global symmetries where $n=2$; (2) the class with $Z_2$ or higher discrete global symmetries where $n\geq 4$. Below we will focus on $Z_1$ ($n=2$) and $Z_2$ ($n=4$) cases only, as the conclusions on the cases with discrete global symmetries higher than $Z_2$ symmetries remain the same as that for the $Z_2$ case. We also note that the low energy degrees of freedom only depend on the presence or absence of  discrete symmetries, but does not depend on whether the symmetry is unitary or anti-unitary.

For strongly interacting TQCPs, the effect of global symmetries has been analyzed before. The presence of global symmetries does change the universality of these TQCPs as suggested in Ref. \cite{Zhou22}.
On the other hand, for weakly interacting TQCPs between gapped superfluid or superconducting phases, $Z_2$ global symmetries, such as TRS, are found not to change the universal scaling of {\em thermodynamic quantities} near TQCPs or the order of phase transitions \cite{Yang19} in high dimensions $d=2,3$. However, as to be elaborated in this  article, the entropy density and scaling of dynamics such as thermal conductivity are distinct in quantum critical regimes with and without any global symmetries. Practically, they can be robust smoking guns for detecting underlying global symmetries and their impact on relevant degrees of freedom in emergent real fermion dynamics.

\section{Effective field theory and thermodynamic properties}

For concreteness, we use a $Z_2$ TRS as an example to demonstrate the effect of global symmetries. But the results are applicable to discrete symmetries in general, as elaborated above. We will carry out the discussions in the context of TQCPs in superconductors. 
We compare TQCPs with no symmetry ($Z_1$ class) and those with only TRS ($Z_2$ class). 

Following previous works \cite{Yang19,Yang21,Zhou22}, the TQCPs can be described by an effective field theory of real fermions in the superconducting bulk interacting with massive real scalar fields.
The generic low energy effective Hamiltonian for the bulk near TQCPs has the following emergent relativistic form
\begin{equation}\label{Hamiltonian}
\mathcal{H}=\frac12\psi^T\Bigg[c_\psi\sum_{j=1}^d(-i\partial_j)\Gamma_j+\Gamma_0 m\Bigg]\psi+\mathcal{H}',
\end{equation}
\begin{equation}\label{Hint}
\mathcal{H}'=\frac12\big[\pi^2+c_\phi^2(\nabla\phi)^2+M^2\phi^2\big]+g\psi^T\Gamma_0\psi\phi.
\end{equation} 
$\psi$ is the real fermion field in the bulk. It is a two-component ($n=2$) real fermionic field for $Z_1$ class, 
and a four-component ($n=4$) real fermionic field for $Z_2$ class. $\Gamma_i, i=1,...,d$, and $\Gamma_0$ are mutually anti-communting matrices whose details depend on representations \cite{Gamma}. 
$\phi$ is a massive real scalar field representing the Higgs mode in superconductors, $\pi$ is the canonical conjugate field of $\phi$, $M$ is proportional to the condensate amplitude, and $d$ is the spatial dimension. The fermion gap in the bulk is given by mass $m$. A TQCP occurs at $m=0$ when bulk gap closes. The mass operator is consistent with the symmetries of the gapped phases on both sides of the topological transition and can be tuned by, e.g., the chemical potential. Here, we focus on a generic model where fermions only couple to one single most relevant scalar field.
When $c_\phi=c_\psi$, the Hamiltonian further exhibits an emergent Lorentz symmetry.

The effective Hamiltonian is valid in the vicinity of the TQCP where real fermions are almost massless
but $\phi$ field is always massive \cite{Yang19} unless further fine tuned so that the TQCP falls into a conformal fixed point.
Therefore, generically near TQCPs, we can have $m \ll \Lambda_T \ll M$, with $\Lambda_T$ being a running (ultraviolet) scale of the effective theory; $\Lambda_T$ can be simply set to be $T$ for our purpose. 

Let us first focus on the weakly interacting case.
A direct consequence of different degrees of freedom near TQCPs due to different symmetries is different amounts of thermal entropy in the quantum critical regime. The degrees of freedom near a TQCP of $Z_2$ class is twice that of a TQCP of $Z_1$ class. 
Thus, thermal entropy in the quantum critical regime should also be twice for systems with a $Z_2$ symmetry, like TRS, compared with those without any symmetries. 
Near a weakly interacting TQCP, interactions always flow to zero \cite{Yang19}. Therefore, we can approximate the thermal entropy density $S$ in the quantum critical regime using simple thermodynamic relations for noninteracting systems. The free energy density $F$ is given by \cite{Coleman}
\begin{equation}
F\approx-\frac{(2s+1)T}V\sum_{\bf k}\ln(1+e^{-\beta\epsilon_{\bf k}})\approx -(2s+1)\alpha_d\frac{T^{d+1}}{c_\psi^d},
\end{equation}
with $V$ the total volume and $\epsilon_{\bf k}=\sqrt{c_\psi^2k^2+m^2}$.
Therefore, the thermal entropy density is 
\begin{equation}
S=-\frac{\partial F}{\partial T} \approx(2s+1)(d+1)\alpha_d c_\psi^{-d}T^d,
\end{equation}
where $s$ is the total spin, and $\alpha_d=(2\pi)^{-d}\Omega_{d-1}\int dx x^{d-1}\ln(1+e^{-x})$, with $\Omega_{d-1}$ the solid angle for a $(d-1)$-sphere.
Note that the $Z_2$ class with TRS has $s=1/2$, while the $Z_1$ class breaking TRS has $s=0$. For $Z_2$ symmetries other than TRS, $s$ can be viewed as a pseudo-spin.
Thus, although the entropy always scales as $T^d$ regardless of global symmetries, the value of entropy density measured in units of $(T/c_\psi)^d$ is doubled when a $Z_2$ symmetry is present;
so is the specific heat $C_V$ that directly measures the entropy.

\section{Thermal conductivity}

\begin{figure}
\centering
\includegraphics[width=0.7\columnwidth]{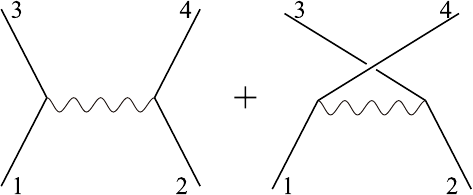}
\caption{\label{scattering} Effective fermion-fermion scattering in the massive scalar field limit. Solid lines represent two-component real fermions with given spin, and wavy lines represent massive scalar fields.}
\end{figure}

Another aspect of different symmetries is the scaling of transport properties such as longitudinal thermal conductivity.
For the weakly interacting case, since the scalar field $\phi$ is too massive to be excited, it is the elastic scattering between fermions that enters thermal conductivity.
$Z_2$ symmetry such as TRS plays an important role in determining the scaling of the scattering amplitude. 
In the following, we consider scattering between two-component real fermions with given spin $(\chi_{+,s_i},\chi_{-,s_i})^T$ (Fig.\ref{scattering})
where the index $i$ denotes energy $\epsilon_i$, momentum ${\bf k_i}$, and spin $s_i$.

For the $Z_1$ class of spinless fermions, we find that the direct and exchange processes in Fig.\ref{scattering} cancel in the leading order $O(\frac{1}{M^2})$ resulting in a subleading term of  $O(\frac{1}{M^4})$ in the limit $M\gg \Lambda_T$,
\begin{equation}
\begin{split}
a(1,2;3,4)\propto &\frac{g^2}{M^4}\Big[(\epsilon_1-\epsilon_4)^2-c_\phi^2({\bf k_1-k_4})^2\\
&-(\epsilon_1-\epsilon_3)^2+c_\phi^2({\bf k_1-k_3})^2\Big].
\end{split}
\end{equation}
Thus, for the $Z_1$ case without any symmetry, $a$ vanishes in the infrared limit when the frequency and momentum approach zero with a scaling dimension $D_s=2$.

In contrast, for the $Z_2$ case of spin-$1/2$ fermions, the direct and exchange processes do not cancel and the term of $O(\frac{1}{M^2})$ is dominating in the same limit, and we have
\begin{equation}
a(1,2;3,4)\propto \frac{g^2}{M^2}\delta_{s_1,-s_2}\delta_{s_3,-s_4}\delta_{s_1,s_3}.
\end{equation}
$a$ in the presence of a global symmetry such as a $Z_2$ (or higher symmetry groups) remains a constant in the infrared limit with a scaling dimension $D_s=0$, in stark contrast to the $Z_1$ case without global symmetries.

The scattering amplitude between real fermions sets the scattering rate between quasiparticles.
This scattering rate can be obtained from Fermi's Golden rule \cite{Coleman, supp}.
Accordingly, the scattering rate has the following scaling behaviors for low energy quasiparticles with $\epsilon_1\sim O(T)$.
For the $Z_1$ case, we have 
\begin{equation}
\frac1{\tau_{sc}}\propto \frac{g^4}{M^8}T^{2d+3}.
\end{equation}
In contrast, for the $Z_2$ case we have
\begin{equation}
\frac1{\tau_{sc}}\propto \frac{g^4}{M^4}T^{2d-1}.
\end{equation}

As a result, longitudinal thermal conductivity in the quantum critical regime should scale differently for systems with and without $Z_2$ symmetries.
Applying the Boltzmann transport equations, we indeed find that the longitudinal thermal conductivity scales differently with temperature in the quantum critical regime for different symmetries. For the $Z_1$ class with no global symmetries, we have
\begin{equation}\label{kappaB}
\kappa\propto T^{-(d+3)}.
\end{equation}
While for the $Z_2$ class, it becomes
\begin{equation}\label{kappaI}
\kappa\propto T^{-(d-1)}.
\end{equation}
These scalings are valid in a typical quantum critical regime with $M \gg T \gg m$.
This is one of the main results of this  article. 
The example of the $Z_1$ class can be two dimensional $p+ip$ chiral superfluids ($d=2$). Meanwhile, $Z_2$ TRS protected topological superconductors can appear in both two and three dimensions ($d=2,3$). 

We would like to emphasize that different scalings of transport quantities and mass renormalization are unique to gapless real fermions \cite{mass}. This robust difference arises from the fact that when no symmetry is present, TQCP dynamics are determined by minimum two-component real fermion fields (i.e., $n=2$).
Two real fermions cannot interact locally and there can be no four-fermion operators without involving gradient operators. 
The most relevant four-fermion interaction operator has the scaling dimension of $D_{4\psi}=d+1$. This is in stark contrast to the case of $n \geq 4$ fermions, where four-fermion local interactions (without gradient operators) typically dominate in the long wavelength limit and the most relevant four-fermion operator has a lower scaling dimension of $D_{4\psi}=d-1$. The distinction between $Z_1$ and other cases is generic and universal, and is independent of the models we have adapted for quantitative discussions.

\section{Surface Majorana states}

Without fine tuning, a topological superconductor typically is not  gapless or quantum critical. However, practically the results above can be easily applied to other gapless Majorana systems. For example, the surface Majorana states of 3D topological superconductors are always gapless and can be viewed as quantum critical from a surface point of view \cite{Yang21}. Therefore, it might be more plausible to examine the role of global symmetries by studying thermal conductivity of surface states. 
There is a subtle difference between bulk and surface states. Since the degrees of freedom of surface states are half of those in the bulk, surface states with only one $Z_2$ global symmetry such as TRS are described by two-component real fermions rather than four-component real fermions in the bulk, mapping the dynamics into the $Z_1$ case. 

For instance, consider a 3D topological superconductor with a global $Z_2$ symmetry (e.g., TRS). The effective Hamiltonian on the $xz$-surface can be written as
\begin{equation}
\mathcal{H}_\text{surf}=\frac12\psi^Tc_\psi[\sigma_z(i\partial_x)-\sigma_x(i\partial_z )]\psi+\mathcal{H}',
\end{equation}
where $\psi=(\chi_{+\uparrow},\chi_{+\downarrow})^T$ is a two-component real fermion on the surface. $\sigma_{x,z}$ are Pauli matrices in the spin space, and $\mathcal{H}'$ is the same as Eq. (\ref{Hint}).
This effectively describes a two-dimensional weakly-interacting massless real system, which can be mapped into a 2D TQCP without any global symmetry, i.e. the $Z_1$ class. 
Thus, the surface transport properties of 3D topological superconductors with a $Z_2$ symmetry should resemble the transport properties of a 2D topological superconductor with no symmetries in the quantum critical regime. Consequently, the surface thermal conductivity should scale as Eq. (\ref{kappaB}) with $d=2$.
Since the surface states remain gapless as long as the bulk gap remains open, no fine-tuning is required to observe such dynamics.

\section{Strong coupling limit}

Next, let us briefly comment on strongly interacting cases. 
Details of thermal transport in this limit remain a challenging topic.
Here we focus on transport properties dictated by strong coupling fixed points. The strong coupling limit corresponds to massless scalar fields or $M=0$ limit. The one-loop renormalization group equation for $g$ has the following form
\begin{equation}
\frac{d\tilde g^2}{d\ln(\Lambda/\Lambda_0)}=-\epsilon \tilde g^2+c_d\tilde g^4,
\end{equation}
where $\Lambda$ and $\Lambda_0$ are the running and ultraviolet momentum cutoff, $\epsilon=3-d$, $\tilde g^2=g^2/\Lambda^\epsilon$, and $c_d$ is a numerical factor \cite{Zhou22}.
This suggests an {\em infrared stable strong coupling fixed point} at $M=0$, $\tilde g_c^2=c_d^{-1}\epsilon$ for dimensions $d=3-\epsilon < 3$.
These strong coupling fixed points belong to the Gross-Neveu type and are represented by conformal field theories.
The fixed point of the $Z_1$ class further exhibits supersymmetry in addition to the standard scale-conformal symmetry \cite{Zhou22}.
Nevertheless, in both $Z_1$ and $Z_2$ classes, dynamics are manifestly scale invariant, differing from the weakly coupling cases as the scalar field is now massless.

At finite $T$, the scalar fields develop a mass gap of order of $f(\tilde{g}_c^2) T$ where $f(\tilde{g}_c^2=c_d^{-1}\epsilon)$ is a universal function that only depends on $\epsilon$.
In the standard $\epsilon$-expansion, one obtains that $f(\epsilon)\propto \epsilon$ \cite{supp}.
Following general considerations, the   scattering rate has the following generic scaling form, for both $Z_1$ and $Z_2$ type {\em strong coupling fixed points},
$\frac{\hbar}{\tau_{sc}}\propto\Lambda^z h(\tilde{g} (\Lambda), \frac{\Lambda^z}{T})$, where $\Lambda$ is the ultraviolet scale of an effective field theory, $\tilde{g}(\Lambda)$ is the dimensionless coupling constant for a running scale $\Lambda$ and $h(x,y)$ is 
a dimensionless function.
$z=1$ is the dynamic scaling exponent in our case.

The key idea here is that at strong coupling {\em fixed points}, $\tilde{g}(\Lambda)=\tilde{g}_c \sim \epsilon$ is independent of the running scale $\Lambda$.
So by simply setting the running scale so that $\Lambda^z=T$, we find that the   scattering rate $1/\tau_{sc}$ is naturally related to the Planckian time scale $\hbar/T$ at fixed points, regardless of symmetry. Indeed, we find that in $d=3-\epsilon$ (i.e. slightly below 3D) \cite{supp}
\begin{eqnarray}
\frac{\hbar}{\tau_{sc}} =T h(\tilde{g}_c, 1)\propto \epsilon T.
\end{eqnarray}
Consequently, unlike in the weakly coupling limit where we have found distinctly different scaling properties, longitudinal thermal conductivity at strong coupling fixed points shall always scale as, for both $Z_1$ and $Z_2$ classes,
\begin{eqnarray}
\kappa \propto T^{d-1}.
\end{eqnarray}
So in this limit, scaling behaviors of thermal transport $\kappa$ turn out to be the same with or without $Z_2$ global symmetries.

\section{Discussions and Conclusion}
So far, we have mainly focused on the effects of discrete global symmetries without considering any addiotional continuous symmetries, and we can exclusively restrict ourselves to the field theories of  dynamic exponents $z=1$. 
However, it is straightforward to generalize to gapless fermions with other dynamic exponents such as $z=2$ Lifshitz Majorana fields \cite{Yang21} that form natural representations for TQCPs with potential nodal structures or additional continuous symmetries.  
So before closing, we list the results for $z=2$ fields in the {\it weak coupling limit} here as a brief reference. We have 
\begin{equation}
\kappa \propto T^{-\frac{d}{2}}
\end{equation}
for the $Z_1$ class without any discrete global symmetries, and
\begin{equation}
\kappa \propto T^{-(\frac{d}{2}-2)}
\end{equation}
for the class with $Z_2$ or higher discrete global symmetries. Like in the previous discussions, a discrete global symmetry again plays a paramount role in thermal transport and sets the scaling properties. 
 
In conclusion, we have examined the effect of discrete global symmetries on interacting gapless Majorana fermions in superfluids and superconductors. We find that discrete global symmetries can set the values of entropy density and scaling properties of transport quantities such as thermal conductivity. Global symmetries determine degrees of freedom at low energies, which not only directly affect entropy but also put stringent constraints on effective fermion-fermion interactions. In this  article, we have used TRS as an example to discuss these effects, but the results can be generalized to other global symmetries such as parity and crystalline symmetries. We also note that the scalings of thermal conductivity are applicable to gapless surface Majorana systems that are accessible to experiments.

One of the applications of the results obtained in this  article is to utilize them as an approach to detect Majorana fermions in the presence or absence of global symmetries as thermal transport properties are distinctly different. 
Generally, global symmetries such as TRS cannot be easily detected directly in experiments. The measurement on thermal conductivity can offer information about possible global symmetries present in interacting Majorana fermions.

\begin{acknowledgments}
F.Y. is supported by Chinese International Postdoctoral Exchange Fellowship Program (Talent-introduction Program) and Shuimu Tsinghua Scholar Program at Tsinghua University.
F.Z. is supported by an NSERC (Canada) Discovery Grant under the contract RGPIN-2020-07070 and a grant from the University of British Columbia.
\end{acknowledgments}

\appendix

\section{Computing thermal conductivity using Fermi's golden rule and Boltzmann equation}
The scattering rate is given by Fermi's golden rule
\begin{equation}\label{FGR}
\begin{split}
\frac1{\tau_{sc}}=\frac{2\pi}{\hbar}\sum_{\substack{{\bf k_2, k_3, k_4}\\s_2,s_3,s_4}}&\Big\{ |\tilde a(1,2;3,4)|^2\delta(\epsilon_1+\epsilon_2-\epsilon_3-\epsilon_4)\\
&\times(2\pi\hbar)^d\delta({\bf k_1+k_2-k_3-k_4})\\
&\times[f_2(1-f_3)(1-f_4)+(1-f_2)f_3f_4]\Big\},
\end{split}
\end{equation}
where $f_i=1/(e^{\epsilon_i}+1)$ and $\epsilon_i=\sqrt{c_\psi^2{\bf k_i}^2+m^2}$ are the quasiparticle distribution function and energy, and $\tilde a(1,2;3,4)$ is the scattering amplitude between quasiparticles with coherence factors further included \cite{Schrieffer}.

The scattering amplitude of quasiparticles $\tilde a(1,2;3,4)$ differs from that of real fermions $a(1,2;3,4)$ due to the presence of coherence factors.
For the $Z_1$ class, $\tilde a$ is
\begin{equation}
\begin{split}
\tilde a\propto \frac{g^2}{M^4}&\bigg\{\Big[\big((\epsilon_1-\epsilon_4)^2-c_\phi^2({\bf k_1-k_4})^2\big)\\
&\times(u^*_{\bf k_1}u_{\bf k_4}-v_{\bf k_1}v_{\bf k_4}^*)(u^*_{\bf k_2}u_{\bf k_3}-v_{\bf k_2}v_{\bf k_3}^*)\Big]\\
&-\Big[\big((\epsilon_1-\epsilon_3)^2-c_\phi^2({\bf k_1-k_3})^2\big)\\
&\times(u^*_{\bf k_1}u_{\bf k_3}-v_{\bf k_1}v_{\bf k_3}^*)(u^*_{\bf k_2}u_{\bf k_4}-v_{\bf k_2}v_{\bf k_4}^*)\Big]\bigg\}.
\end{split}
\end{equation}
At the TQCP, we have $u_{\bf k}=1/\sqrt2$ and $v_{\bf k}=\Delta_{\bf k}/(\sqrt2|\Delta_{\bf k}|)$, where we have chosen $u_{\bf k}$ to be real and positive by convention. $\Delta_{\bf k}$ is the superconducting order parameter. For the $Z_1$ example considered in this  article, we choose a time-reversal breaking $p+ip$ superconductor with $\Delta_{\bf k}=c_\psi(k_x+ik_y)$.

For the $Z_2$ class, we have
\begin{equation}
\begin{split}
\tilde a\propto &\frac {g^2}{M^2}(u_{\bf k_1}^\dagger u_{\bf k_3}-v_{\bf k_1}v^\dagger_{\bf k_3})_{s_1s_3}(u_{\bf k_2}^\dagger u_{\bf k_4}-v_{\bf k_2}v^\dagger_{\bf k_4})_{s_2s_4}\\
&\times\delta_{s_1,-s_2}\delta_{s_1,s_3}\delta_{s_2,s_4}.
\end{split}
\end{equation}
Here $u_{\bf k}$, $v_{\bf k}$ are $2\times2$ matrices. At the TQCP, we have $u_{{\bf k}\sigma\sigma'}=\delta_{\sigma\sigma'}/\sqrt2$, and $v_{{\bf k}\sigma\sigma'}=\Delta_{{\bf k}\sigma\sigma'}/\sqrt{2(\Delta_{{\bf k}}\Delta_{{\bf k}}^\dagger)_{\sigma\sigma}}$. Here, we choose the time-reversal invariant Balian-Werthamer (BW) phase of Helium 3 as an example, whose order parameter is given by $\Delta_{\bf k}=c_\psi i{ \bm {\sigma}}\cdot{\bf k}\sigma_y$.

The thermal conductivity can be computed using the Boltzmann equation
\begin{equation}
\partial_t f'+(\nabla_{\bf k}\epsilon_{\bf k})\cdot({\nabla_{\bf r}} f')-(\nabla_{\bf r}\epsilon_{\bf k})\cdot(\nabla_{\bf k}f')=-\frac{f'-f}{\tau_{sc}},
\end{equation}
where $\nabla_{\bf r}$ and $\nabla_{\bf k}$ are gradients in spatial and momentum spaces, respectively, and $f'$ and $f$ are quasiparticle distribution functions away and at equilibrium. In the limit $m\to0$, we have
\begin{equation}
\kappa=(2s+1)\tau_{sc} T^dc_\psi^{2-d} I_d,
\end{equation}
where $I_d=\int\frac{d\Omega_{d-1}}{(2\pi)^d}\cos^2\theta\int dx x^{d+1}\frac{e^x}{(1+e^x)^2}$.

\section{Fermion mass renormalization in the weak coupling limit}

\begin{figure}[h]
\centering
\includegraphics{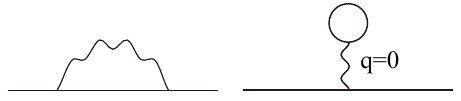}
\caption{\label{mass} One-loop diagram for the fermion propagator.}
\end{figure}

Global symmetries affect the fermion mass renormalization in the weak coupling limit, which is given by the diagrams in Fig. \ref{mass}.
The two diagrams differ by a factor of $-(2s+1)$ as a result of the fermion loop in the second one. For $Z_1$ class of spinless fermions ($s=0$), the two diagrams exactly cancel each other in the order of $1/M^2$ in the large mass limit $M\gg \Lambda_0$. Thus, the fermion mass renormalization is given by
\begin{equation}
\delta m^{(B)}\approx\frac{g^2c_\psi\Lambda_0^3}{18\pi^2M^4}\left(1+\frac{2c_\phi^2}{c_\psi^2}\right)m.
\end{equation}
However, for $Z_2$ class of spin-1/2 fermions, these two diagrams no longer cancel in the order of $1/M^2$, and the fermion mass renormalization becomes
\begin{equation}
\delta m^{(I)}\approx(2s)\frac{g^2\Lambda_0}{2\pi^2 M^2c_\psi} m.
\end{equation}

\section{Strong coupling limit}
We discuss the detailed structure of the imaginary part of fermion self-energy near strong coupling fixed point. For simplicity, we focus on the case where $c_\phi=c_\psi=1$. 
The scalar field acquires a finite mass at $T> 0$ and $d<3$. Let $\epsilon=3-d$, we have
\begin{equation}
M^2=(2s+1)g^2T^{d-1}\int \frac{d^d x}{(2\pi)^d}\frac1{x(e^x+1)}.
\end{equation}
As $g_c^2\sim \epsilon T^{3-d}$, we have $M^2\sim\epsilon T^2$.

The imaginary part of the on-shell retarded self-energy can be written as
\begin{equation}
\text{Im}\Sigma^R(\epsilon_{\bf k},{\bf k})=\text{Im}\Sigma_0^R(\epsilon_{\bf k},{\bf k})I+\text{Im}\Sigma_j^R(\epsilon_{\bf k},{\bf k})\Gamma_j,
\end{equation}
where $I$ is the identity matrix, the repeated index $j$ is summed over from 1 to $d$.

We have
\begin{equation}
\begin{split}
\text{Im}\Sigma_0^R(\epsilon_{\bf k},{\bf k})=&-\pi g^2\int \frac{d^dq}{(2\pi)^d}\frac1{4\xi_{\bf q}}\Big\{(1-f_{\bf k-q}+n_{\bf q})\\
&\times\big[\delta(\epsilon_{\bf k}-\xi_{\bf q}-\epsilon_{\bf k-q})+\delta(\epsilon_{\bf k}+\xi_{\bf q}+\epsilon_{\bf k-q})\big]\\
&+(f_{\bf k-q}+n_{\bf q})\big[\delta(\epsilon_{\bf k}-\xi_{\bf q}+\epsilon_{\bf k-q})\\
&\qquad+\delta(\epsilon_{\bf k}+\xi_{\bf q}-\epsilon_{\bf k-q})\big]\Big\},
\end{split}
\end{equation}
and 
\begin{equation}
\begin{split}
\text{Im} \Sigma_j^R(\epsilon_{\bf k},{\bf k})=&\pi g^2\int \frac{d^dq}{(2\pi)^d}\frac{k_j-q_j}{4\xi_{\bf q}\epsilon_{\bf k-q}}\Big\{(1-f_{\bf k-q}+n_{\bf q})\\
&\times\big[\delta(\epsilon_{\bf k}-\xi_{\bf q}-\epsilon_{\bf k-q})-\delta(\epsilon_{\bf k}+\xi_{\bf q}+\epsilon_{\bf k-q})\big]\\
&+(f_{\bf k-q}+n_{\bf q})\big[-\delta(\epsilon_{\bf k}-\xi_{\bf q}+\epsilon_{\bf k-q})\\
&\qquad+\delta(\epsilon_{\bf k}+\xi_{\bf q}-\epsilon_{\bf k-q})\big]\Big\}.
\end{split}
\end{equation}
Here $\epsilon_{\bf k-q}= |{\bf k-q}|$, $\xi_{\bf q}=\sqrt{q^2+M^2}$, $f_{\bf k-q}=1/(e^{\epsilon_{\bf k-q}}+1)$, and $n_{\bf q}=1/(e^{\xi_{\bf q}}-1)$. For $M>0$, only $\delta(\epsilon_{\bf k}-\xi_{\bf q}+\epsilon_{\bf k-q})$ can be satisfied. This puts a constraint on the angle between ${\bf k}$ and ${\bf q}$ for each given $\xi_{\bf q}$.

The $\delta$-function can be rewritten as
\begin{equation}
\begin{split}
\delta(\epsilon_{\bf k}-\xi_{\bf q}+\epsilon_{\bf k-q})=&\delta(\cos\theta-\cos\theta_0)\\
&\times\Bigg(\frac{\sqrt{1+M^2/q^2}}{k}-\frac1q\Bigg)\Theta_1\Theta_2\Theta_3,
\end{split}
\end{equation}
where $\theta$ is the angle between ${\bf k}$ and ${\bf q}$,
\begin{equation}
\cos\theta_0=\sqrt{1+\frac{M^2}{q^2}}-\frac{M^2}{2kq},
\end{equation}
\begin{equation}
\Theta_1=\Theta(\sqrt{q^2+M^2}-k),
\end{equation}
\begin{equation}
\Theta_2=\Theta\left(q-k+\frac{M^2}{4k}\right),
\end{equation}
\begin{equation}
\Theta_3=\Theta\left(k^2-\frac{M^4}{8q^2+4M^2+8q\sqrt{q^2+M^2}}\right),
\end{equation}
and $\Theta(\cdot)$ is the Heaviside step function.

Thus, we have
\begin{equation}
\begin{split}
\text{Im}\Sigma^R(\epsilon_{\bf k}, {\bf k})=&-\pi g^2\int \frac{d^dq}{(2\pi)^d}\Bigg\{\Bigg(\frac1{4\xi_{\bf q}}I+\frac{k_j-q_j}{4\xi_{\bf q}\epsilon_{\bf k-q}}\Gamma_j\Bigg)\\
&\times\Bigg[(f_{\bf k-q}+n_{\bf q})\Bigg(\frac{\sqrt{1+M^2/q^2}}{k}-\frac1q\Bigg)\\
&\qquad\times \delta(\cos\theta-\cos\theta_0)\Theta_1\Theta_2\Theta_3\Bigg]\Bigg\},
\end{split}
\end{equation}
where the summation over index $j$ is assumed.
For $d=3-\epsilon$, we can first perform the angular integration, this restricts the direction of ${\bf q}$ for each given ${\bf k}$. The step functions puts a constraint on the minimum value of $q$. Although $\text{Im}\Sigma^R(\epsilon_{\bf k}, {\bf k})$ is anisotropic, all its matrix elements have the same scaling in $T$.
For $\epsilon_{\bf k}\approx M/2$, the step functions can be satisfied for $q_\text{min}\to 0$. In this case, each matrix element of $\text{Im}\Sigma^R(\epsilon_{\bf k}, {\bf k})$ is proportional to
\begin{equation}
g^2 T^{d-2}\ln\epsilon\propto T \epsilon  \ln\epsilon, \qquad \epsilon_{\bf k}\approx \frac M2.
\end{equation}
For $\epsilon_{\bf k}\gg M/2$, the step functions require that $q_\text{min}\gtrsim k\gg M$. For $\epsilon_{\bf k}\ll M/2$, the step functions require $q_\text{min}\gg M^2/k\gg M$. In both cases, each matrix element of $\text{Im}\Sigma^R(\epsilon_{\bf k}, {\bf k})$ is proportional to
\begin{equation}
g^2 T^{d-2}\propto \epsilon T, \qquad \epsilon_{\bf k}\ll \frac M2 \quad\text{or}\quad \epsilon_{\bf k}\gg \frac M2.
\end{equation}

For the typical case, the fermion energy $\epsilon_{\bf k}$ is of the order $O(T)$, which is much larger than $M/2\sim O(\epsilon T)$. 
The scattering rate between quasiparticles is determined by $\text{Im}\Sigma^R(\epsilon_{\bf k},{\bf k})$. 
Thus, the scattering rate has the following scaling
\begin{equation}
\frac1{\tau_{sc}}\propto \epsilon T.
\end{equation}


\begin{thebibliography}{References}
\bibitem{Gamma} For superconductors with no symmetries, we have $\psi=(\chi_+,\chi_-)^T$ and it is convenient to choose $\Gamma_1=\tau_z$, $\Gamma_2=\tau_x$, $\Gamma_0=\tau_y$.
For topological superconductors with only TRS, we have $\psi=(\chi_{+\uparrow},\chi_{+\downarrow},\chi_{-\uparrow},\chi_{-\downarrow})^T$ and we can choose $\Gamma_1=-\tau_z\sigma_z$, $\Gamma_2=\tau_x\sigma_0$, $\Gamma_3=\tau_z\sigma_x$, $\Gamma_0=\tau_y\sigma_0$. $\tau_\alpha$ and $\sigma_\alpha$ ($\alpha=x,y,z$) are Pauli matrices in the $(\chi_+,\chi_-)$ and spin subspaces, respectively, with $\sigma_0$ being the identity matrix. In this basis, TRS can be written as $\mathcal{T}=\tau_z(i\sigma_y)K$ with $K$ being complex conjugation.
\bibitem{mass}Global symmetries also affect the fermion mass renormalization. For $Z_1$ class, in the large mass limit the fermion mass renormalization is given by
$\delta m\approx\frac{g^2c_\psi\Lambda_0^3}{18\pi^2M^4}\left(1+\frac{2c_\phi^2}{c_\psi^2}\right)m$.
However, for $Z_2$ class, the fermion mass renormalization becomes
$\delta m\approx(2s)\frac{g^2\Lambda_0}{2\pi^2 M^2c_\psi} m$. See Appendix for detail.
The expression for $\delta \mu$ below Eq. (30) in Ref. \cite{Yang19} is only valid for TRI cases.
\end{thebibliography}
\end{document}